\documentclass[sigconf]{acmart}

\usepackage{algorithmic}
\usepackage{graphicx}
\usepackage{textcomp}
\usepackage{xcolor}
\usepackage{fancyhdr}
\usepackage{hyperref}
\usepackage{soul}

\newcommand{\ardhi}[1]{{\color{black}#1}}

\usepackage{listings}
\usepackage{xcolor}
\definecolor{codegreen}{rgb}{0,0.6,0}
\definecolor{codegray}{rgb}{0.5,0.5,0.5}
\definecolor{codepurple}{rgb}{0.58,0,0.82}
\definecolor{backcolour}{rgb}{0.95,0.95,0.92}

\lstdefinestyle{mystyle}{
    backgroundcolor=\color{backcolour},   
    commentstyle=\color{codegreen},
    keywordstyle=\color{magenta},
    numberstyle=\tiny\color{codegray},
    stringstyle=\color{codepurple},
    basicstyle=\ttfamily\scriptsize,
    breakatwhitespace=false,         
    breaklines=true,                 
    captionpos=b,                    
    keepspaces=true,                 
    numbers=left,                    
    numbersep=5pt,                  
    showspaces=false,                
    showstringspaces=false,
    showtabs=false,                  
    tabsize=2
}
\usepackage{pifont}
\newcommand{\cmark}{\ding{51}}%
\newcommand{\xmark}{\ding{55}}%

\usepackage{awesomebox}
\usepackage{lipsum}
\usepackage{tikz}
\usepackage{multirow}
\newcommand*\circled[1]{\tikz[baseline=(char.base)]{\node[shape=circle,draw,inner sep=1pt] (char) {#1};}}

\AtBeginDocument{%
  }

\setcopyright{acmcopyright}\acmConference[PACT '24]{Proceedings of the 2024 International Conference on Parallel Architectures and Compilation Techniques}{October 13--16, 2024}{Long Beach, CA, USA}
\acmBooktitle{Proceedings of the 2024 International Conference on Parallel Architectures and Compilation Techniques (PACT '24), October 13--16, 2024, Long Beach, CA, USA}
\copyrightyear{2024}

\begin{document}

\title{BoostCom: Towards Efficient Universal Fully Homomorphic Encryption by Boosting the Word-wise Comparisons}

\author{Ardhi Wiratama Baskara Yudha}
\email{yudha@ucf.edu}
\affiliation{\institution{University of Central Florida}}

\author{Jiaqi Xue}
\email{jiaqi.xue@ucf.edu}
\affiliation{\institution{University of Central Florida}}

\author{Qian Lou}
\email{qian.lou@ucf.edu}
\affiliation{\institution{University of Central Florida}}

\author{Huiyang Zhou}
\email{hzhou@ncsu.edu}
\affiliation{\institution{North Carolina State University}}

\author{Yan Solihin}
\email{yan.solihin@ucf.edu}
\affiliation{\institution{University of Central Florida}}

\begin{abstract}

Fully Homomorphic Encryption (FHE) allows for the execution of computations on encrypted data without the need to decrypt it first, offering significant potential for privacy-preserving computational operations. Emerging arithmetic-based FHE schemes (ar-FHE), like BGV, demonstrate even better performance in word-wise comparison operations over non-arithmetic FHE (na-FHE) schemes, such as TFHE, especially for basic tasks like comparing values, finding maximums, and minimums. This shows the universality of ar-FHE in effectively handling both arithmetic and non-arithmetic operations without the expensive conversion between arithmetic and non-arithmetic FHEs. We refer to universal arithmetic Fully Homomorphic Encryption as uFHE. The arithmetic operations in uFHE remain consistent with those in the original arithmetic FHE, which have seen significant acceleration. However, its non-arithmetic comparison operations differ, are slow, and have not been as thoroughly studied or accelerated. In this paper, 
we introduce BoostCom, a scheme designed to speed up word-wise comparison operations, enhancing the efficiency of uFHE systems. BoostCom involves a multi-prong optimizations including infrastructure acceleration (Multi-level heterogeneous parallelization and GPU-related improvements), and algorithm-aware optimizations (slot compaction, non-blocking comparison semantic). Together, BoostCom achieves an end-to-end performance improvement of more than an order of magnitude (11.1 $\times$ faster) compared to the state-of-the-art CPU-based uFHE systems, across various FHE parameters and tasks.


\end{abstract}

\maketitle

\section{Introduction}

\begin{figure}
  \centering 
  \includegraphics[width=0.8\linewidth]{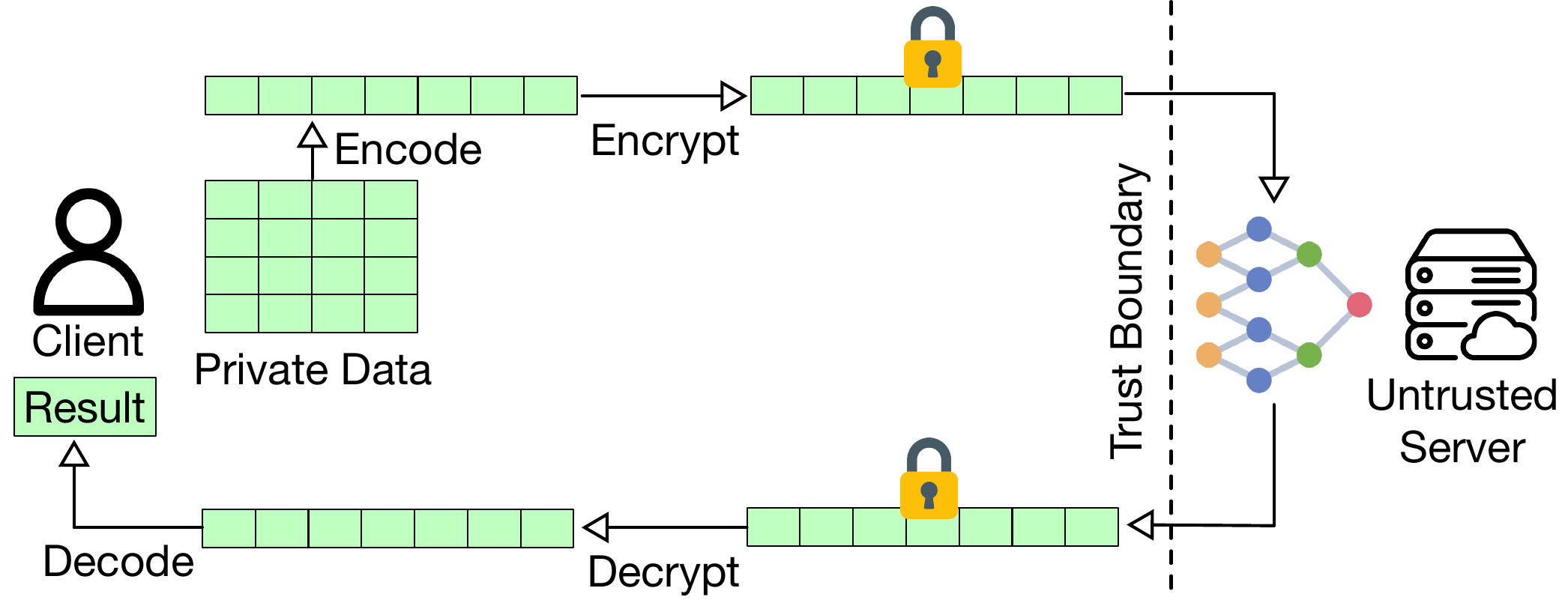}
  \caption{Performing computations on encrypted data transferred to an untrusted server using FHE.}
  \label{fig:fhe_worklflow}
\end{figure}

There has been a surge of interest from the industry in Fully Homomorphic Encryption (FHE)~\cite{gentry2009fully} acceleration recently~\cite{sealcrypto, aharoni2022advanced, IntelHEXL} as FHE may play a pivotal role in facilitating computation on private data in the cloud without disclosing its plaintext. FHE has been cited to be applicable for many types of computation, including machine learning, and big data analytics, on various application domains that include healthcare, finance, genomics research, secure voting systems, and private information retrieval, where it helps maintain stringent privacy regulations~\cite{capeprivacy,DualityTechnologies,Inpher,Zama}. Figure~\ref{fig:fhe_worklflow} illustrates the FHE workflow, which includes client-side encoding, encryption, server-side computation, and subsequent client-end decryption and decoding, which assures client-side data confidentiality even on potentially untrusted servers.

Various FHE schemes have emerged over the past decade, including {\em arithmetic FHE} (ar-FHE), such as word-wise BGV \cite{BGV_scheme} and CKKS \cite{CKKS_scheme}, and {\em non-arithmetic FHE} (na-FHE), such as bit-wise TFHE \cite{TFHE_scheme}. Originally, ar-FHEs were adept at performing arithmetic operations like multiplications and additions, while na-FHEs were primarily used for bit-wise comparison operations. Although na-FHEs excel in bitwise comparisons, they show slower performance in conducting arithmetic operations on integers. In contrast, \textit{the ar-FHE scheme BGV \cite{BGV_comparison} has been upgraded recently with new word-wise comparisons}, such that it not only efficiently handles integer arithmetic operations but also supports batched word-wise comparisons, outperforming na-FHEs in speed. \ardhi{This advancement positions the BGV scheme as a solution for {\em both} arithmetic and non-arithmetic comparisons, a combined capability we henceforth define as {\em universal FHE} (uFHE)}. In contrast, CKKS-based polynomial approximation for non-linear operations still suffers from a precision reduction since each operation affects the fractional value of the ciphertext \cite{Jiang-arxiv22}. Furthermore, comparisons within the CKKS framework, when involving approximated polynomials, lead to non-negligible errors \cite{CKKS-Comparison}\cite{ckks-comparison2}.

Nevertheless, the uFHE scheme based on the new BGV~\cite{BGV_comparison} is not without its limitations, particularly the sluggish and complex comparison operation. A comparison operation compares pairs of encrypted data to generate an encrypted result that indicates whether they are equivalent, less than, or greater than. To execute a single comparison, it requires $3p-5$ non-scalar multiplications along with additions, rotations, and scalar multiplications, with $p$ denoting the plaintext modulus \cite{Tan_scheme}, with typical values reaching up to the tens. Despite the costs, a variety of applications, including scientific computations and machine learning, depend heavily on this comparison operation.

Recognizing that a comparison operation may create a performance bottleneck in new BGV-based uFHE, there has been an effort to rely on an algorithmic approach to accelerate it~\cite{BGV_comparison}. The algorithmic approach reduces the comparison complexity to $2p-6$ (Bivariate case) and $\sqrt{p-3}+\mathcal{O}(\log{p})$ (Univariate case)~\footnote{\ardhi{Bivariate and Univariate are two different algorithms used to perform comparison operations}.}. Although an algorithmic approach is valuable, we are of the view that it alone may not be adequate to meet the requirements of high performance. Proposals have been made to switch between FHE schemes like TFHE-BGV~\cite{Chimera} and TFHE-CKKS~\cite{lu2021pegasus}. However, these transitions are still costly, with over $70\times$ the latency of BGV~\cite{BGV_comparison}. 

Therefore, in this paper, we propose an infrastructure acceleration approach (called {\em BoostCom}), where we offload comparison to the Graphics Processing Unit (GPU) and apply various optimizations. We note that infrastructure acceleration approach has been pursued successfully for various other operations such as encryption, decryption, multiplication, bootstrapping, and other power-of-two polynomial ring operations~\cite{HE-Booster, 100x, HE_on_gpu, CARM, TensorFHE}, including on FPGA~\cite{HEAX, FPGA_HE}, and ASIC~\cite{F1, CraterLake, BTS}. However, they have all neglected the comparison operation, which is the focus of this paper. 

However, a single comparison may be up to multiple orders of magnitude slower than multiplication. Hence, accelerating comparison is challenging, requiring us to use \ardhi{several strategies} including heterogeneous CPU/GPU parallelization, slot compaction, non-blocking comparison semantics, branch removal, and layout optimization, as detailed below. 

  \label{table:operation_latency_comparison}

First, we introduce a strategy for heterogeneous parallelization, wherein multicore CPUs manage parallelization at a higher, digit-level, while GPUs handle the parallelization of fundamental FHE operations at a more granular, polynomial level. This design approach is inspired by the crucial insight that, although the bulk of BGV comparison operations can be parallelized, the parallelism granularity of certain inner operations is insufficient to outweigh the overheads associated with memory copying, memory allocation, and kernel launching. Consequently, transferring these tasks to GPUs might not result in a net gain in performance. By adopting a heterogeneous, multi-level parallelism strategy, we enable CPUs and GPUs to collaborate effectively, thereby enhancing the efficiency of BGV comparison operations. 

Second, we propose multiple GPU-related optimizations for primitive polynomial operations in word-wise comparisons. Profiling detailed in Section~\ref{sec:bottleneck_analysis} reveals that the execution time of word-wise comparison operations is mainly spent on three components: BluesteinNTT, BluesteinFFT, and Element-wise operations. These components are highly parallelizable, suggesting potential efficiency gains by offloading them to GPU. The optimizations include branch removal to increase the parallelism of BluesteinNTT, plan reuse to reduce the computational loads for BluesteinFFT, and memory layout transformation for efficient element-wise operations.


Third, we introduce two algorithmic enhancements in addition to heterogeneous parallelism and GPU optimizations. (I) Slot compaction. uFHE's comparison mechanism leverages SIMD ciphertext batching, allowing the comparison of two vectors through a single ciphertext comparison, given that each vector of size \textit{a} can be encoded into one ciphertext. This is possible when the number of slots (\textit{b}) in the ciphertext exceeds \textit{a}, highlighting the significance of slot utilization ($\frac{a}{b}$) for comparison efficiency. A critical observation is that in prevalent workloads (e.g., machine learning), comparisons typically follow arithmetic operations, leading to low slot utilization in ciphertexts awaiting comparison. This scenario presents an opportunity for slot compaction, enhancing efficiency by increasing slot utilization before comparison. To capitalize on this, we introduce a slot manager designed to track and optimize slot utilization within a ciphertext. This strategy facilitates slot compaction, thereby reducing memory consumption and boosting performance.(ii) non-blocking semantic. We propose \emph{non-blocking semantic} for comparison that allows the overlap of comparison with other computations. The semantic allows comparison to be executed on another CPU thread while the main CPU thread continues executing the next code segment concurrently until the main thread needs to use the result of the comparison. To increase the distance until the use of the result, we perform code straightlining.




We implemented the optimizations on a real-world library (HElib) which enables us to evaluate end-to-end performance (instead of operation-wise evaluation in many prior studies) reliably. \ardhi{Our optimizations do not negatively affect the noise budget, as no additional homomorphic operations were added; they merely enhance GPU efficiency}. We evaluate several applications including sorting, finding minimum elements, multi-layer perceptron (MLP), image re-colorizing, and a private query. Our evaluation shows that the proposed acceleration is effective in boosting the performance of the comparison and the application that uses it. Across the five benchmarks, it achieves {\em end-to-end} geometric mean speedup of $11.1\times$ (up to 26.7$\times$), over an industry-standard FHE library running on 16-core CPUs. \ardhi{BoostCom significantly outperforms HE-Booster \cite{HE-Booster}, a state-of-the-art GPU accelerator for BGV scheme that is also implemented in HElib, by 553\%. Our optimization }

To summarize, this paper makes the following contributions:
\begin{enumerate}
  \item We proposed a multi-level heterogeneous parallelism method as an infrastructure acceleration for comparison in the uFHE scheme.
  \item We proposed multi-prong GPU-related optimizations for accelerating uFHE comparison, including branch-removal, as well as plan reuse and layout optimization. These optimizations are incorporated into a new library called cuHELIB, which builds upon HElib by leveraging GPU technology.
  \item We present new uFHE comparison algorithms featuring slot compaction for ciphertext comparison to lower memory usage and non-blocking comparison to reduce computational dependencies, thereby increasing throughput.
  \item We conducted a comprehensive evaluation of our scheme, considering end-to-end measurements that include CPU-GPU memory copy, kernel launches, and synchronization on five important applications. This approach provides a more holistic assessment compared to extrapolating from operation-wise measurements.
\end{enumerate}

The remainder of the paper is organized as follows. Section 2 discusses the related work,  Section 3 presents the background, Section 4 analyzes the performance bottlenecks of comparison operations in BGV, Section 5 discusses the design of BoostCom and our proposed optimizations, Section 6 presents our experimental methodology, Section 7 discusses our results, and Section 8 concludes.
\section{Background}

\subsection{Word-wise Universal FHE Scheme}
\label{subsec:comparison_algorithm}
Originally, ar-FHEs (arithmetic FHEs) including BGV were adept at performing arithmetic operations
like multiplications and additions. \ardhi{Recently, the ar-FHE scheme BGV \cite{BGV_comparison} has been upgraded with new word-wise comparisons} (arithmetic operations are still the same with prior BGV), such that it not only efficiently handles integer arithmetic operations but also supports batched word-wise comparisons, outperforming na-FHEs in speed. This advancement positions the BGV scheme as a universal FHE solution (uFHE) for both arithmetic and non-arithmetic comparisons. 

Other uFHE methods have been made to switch between FHE schemes like TFHE-BGV~\cite{Chimera} and TFHE-CKKS~\cite{lu2021pegasus}. However, these transitions are still costly, with over 70x latency compared to BGV~\cite{BGV_comparison}.  Thus, the uFHE based on new upgraded BGV~\cite{BGV_comparison} is the-start-of-the-art. However, the current computational bottleneck of uFHEs, particularly the BGV comparison operation, is limited to running on a single CPU.

\begin{table}[htbp]
\small	
  \caption{Parameters used in BGV and comparison operation.}
  \label{tab:notations}
\centering
  {
  \scriptsize
  \begin{tabular}{|c|l|}
    \hline
    Parameter & Description\\
    \hline \hline 
    $p$ & Plaintext coefficient modulus. \\
    $m$ & The order of the cyclotomic ring. \\
    $N$ & The degree of the cyclotomic polynomial.\\
    $Q$ & The product of (prime) moduli: $Q=\prod_{i=0}^L q_i$.\\
    $L$ & Maximum (multiplicative) level.\\
    $\lambda$ & Security level of a given BGV instance.\\
    $\omega$ & Root of unity of twiddle factor for NTT. \\
    $d$ & The dimension of a vector space over a finite field. \\
    $l$ & The length of vectors to be compared. \\
    \hline
  \end{tabular}}
\end{table}

\noindent\textbf{Basics and arithmetic ops. of uFHE-based BGV.} The uFHE-based BGV scheme is a lattice-based encryption based on Ring Learning with Errors (RLWE) problem~\cite{BGV_scheme}. RLWE is a challenging mathematical problem in lattice-based encryption that creates a foundation for developing safe encryption schemes. Table \ref{tab:notations} shows the essential BGV parameters. 
Key parameters include $p$, $m$, and $N$. $p$ defines the plaintext modulus; a higher $p$ enlarges the plaintext space but slows down comparisons. The roles of $m$ and $N$ will be outlined later.

In the BGV scheme, a plaintext is encoded into a polynomial and encrypted to form a ciphertext polynomial. Computation can be performed on the ciphertext, yielding a result also in ciphertext form, which requires decryption to obtain the plaintext. Both plaintext and ciphertext polynomials reside in the same ring with different coefficient moduli, where the ciphertext modulus is significantly larger than the plaintext modulus. The ciphertext polynomial ring ($R_Q$) in the BGV scheme is $C=R_{Q}\times R_{Q}$, where $R_{Q}=\mathbb{Z}_{Q}[x]/(\Phi_m(x))$, and $\Phi_m(x)$ is the $m^{th}$ cyclotomic polynomial with a degree of $N$. The relationship between $m$ and $N$ is determined by the Euler totient function $\varphi$, i.e., $N = \varphi{(m)}$. While prior works use a power-of-two $N$ for simplicity, non-power-of-two $N$ is suggested for better performance and higher security flexibility~\cite{BGV_comparison, GentryEuroCrypt}. ${Q}\in\mathbb{Z}$ is the ciphertext coefficient modulus at level $L$, representing the product of several primes ($q_0, q_1, q_2, ..., q_L$) that fit into the native integer data type. The value of $Q$ determines the multiplicative depth, i.e., the most extended sequence of homomorphic multiplications during computation. $Q$ is typically much larger than $p$, influencing the message expansion rate after encryption. The individual primes $q_i$ are part of the modulus chain.


The BGV scheme utilizes SIMD-style processing, storing multiple integers in one ciphertext to optimize operation throughput. Leveraging ring isomorphism of polynomial modulus enables multiple plaintext slots within a ciphertext. Modular arithmetic is used for homomorphic operations including addition, multiplication, and rotation. However, noise introduced during encryption limits operation numbers and requires a large ciphertext modulus ($Q$).

\noindent\textbf{Non-arithmetic Comparison of uFHE-based BGV.}
The state-of-the-art comparison algorithm was proposed  in~\cite{BGV_comparison}. 
It exploits SIMD-style processing such that many comparisons can be performed in parallel, leading to a small amortized comparison latency. A large integer comparison operand is encoded into an element of $\mathbb{F}_{p^d}^l$. Here, $\mathbb{F}_{p^d}^l$ represents a finite field extension of degree $d$ over a prime field with $p$ elements. The encoding process involves decomposing a large integer into an element in this vector space, where the vector space is of dimension $l$. 

To compare two integers $a$ and $b$, first, each integer is {\em decomposed} into multiple slots in the form of $\mathbb{F}_{p^d}$. For example, $a$ is decomposed into $a_0, a_1, \ldots, a_{l-1}$, where $a_i$ occupying the $i$-th slot. For each slot, using the {\em mod extract} step, each number is further split into multiple digits in the form of $\mathbb{F}_{p}$. For example, $a_0$ is split into $a_{00}$ (the first digit in the first slot), $a_{01}$ (the second digit in the first slot), etc.

To perform comparison, the algorithm first extracts digits of encrypted numbers in $\mathbb{F}_p$, then performs equality ($EQ$) and less than ($LT$) functions for each digit using specific equations. The computation of $LT$ and $EQ$ for each digit is independent. The results of the equality and less than functions on the digits are combined through lexicographical order. First, the lexicographical order is computed for each block of $d$ digits, and then the results are combined using a final equation that returns encrypted ``1" when $a<b$ or encrypted ``0" otherwise. The last two steps that involve ciphertext shifting and multiplying with the result from the equality circuit are called \emph{ShiftMul}, whereas the step for performing a summation of the ciphertext is called \emph{ShiftAdd}. The digit comparison steps are expensive due to repeated ciphertext exponentiations with large exponents for $d\times l$ times, while other steps (\emph{Extraction}, \emph{ShiftMul}, and \emph{ShiftAdd}) are faster. The process represents a Bivariate circuit with separated $LT$ and $EQ$ computations, whereas the Univariate circuit combines $LT$ and $EQ$ circuits differently.

\subsection{Efficient Polynomial and NTT}


To handle the large ciphertext modulus $Q$, the BGV scheme uses a Residue Number System (RNS) format, splitting the polynomial into $L+1$ residue polynomials with coefficients under modulo $q_i$, where $q_i$'s are pair-wise coprime integers. RNS allows for efficient multiplication and addition of ciphertext polynomials using current hardware systems.

To accelerate polynomial multiplication, the Number Theoretic Transform (NTT) is used, converting the polynomial to an integer Discrete Fourier Transform (DFT) representation using a twiddle factor $\omega$ that meets specific conditions. For efficient NTT and INTT, radix-2 NTT implementations are applied when $N$ is a power of two, employing Cooley-Tukey (CT) and Gentleman-Sande (GS) algorithms. The ciphertext polynomial is represented as a matrix of polynomial coefficients in integer DFT representation of size $(L+1)\times\varphi(m)$, enabling straightforward element-wise operations for multiplication and addition between polynomials.

The BluesteinNTT algorithm is used for polynomial conversion between coefficient representation and integer DFT representation when $N$ is a non-power of two. 
The algorithm requires two twiddle factors: TF1, the twiddle factors for polynomial ring $m$, and TF2, the twiddle factors for a power of two polynomial ring.  First,  the input polynomial is multiplied element-wise by TF1 to generate a polynomial $C$. Then, the polynomial $C$ is padded with zero to become C\_pad and then multiplied by polynomial D\_pad. D\_pad is a polynomial generated from TF1. Both polynomials C\_pad and D\_pad have length power of two greater than $2m-1$. The polynomial multiplication between them is accelerated by the radix-2 NTT algorithm (CT and GS) that requires TF2. The multiplication result (C\_pad \textbf{x} D\_pad) is then truncated to have length $m$, with the exceeding coefficient being added to the polynomial. The resulting polynomial is then multiplied element-wise by TF1. Finally, the polynomial is filtered to have a length from $m$ to $N$.



\section{Related Works}

\begin{table}
\small	
  \caption{The comparison of BoostCom vs. prior works.}
  \vspace{-0.15in}
  \label{tab:scheme_comparison}
\centering
  {
  \scriptsize
  \begin{tabular}{|c|c|c||c|c|c|}
    \hline
    Name & Scheme & Comparison  & End-to-End & Platform\\
    \hline \hline
    SHARP~\cite{SHARP_ISCA23}         & CKKS & \cmark      & \xmark      & ASIC \\
    \hline
    CraterLake~\cite{CraterLake} & CKKS & \xmark     & \xmark             &  ASIC\\
    \hline
    FxHENN~\cite{FxHENN}                      & CKKS & \xmark    & \xmark         & FPGA \\
    \hline
    TensorFHE~\cite{TensorFHE} & CKKS &\xmark      & \xmark      & GPU\\
    \hline
    HE on GPU~\cite{HE_on_gpu} & BFV &\xmark      & \cmark      & GPU\\
    \hline
    Intel HEXL~\cite{IntelHEXL} & BGV & \xmark    & \cmark       &  CPU\\
    \hline
    HE-Booster~\cite{HE-Booster} & BGV & \xmark    & \xmark       &  GPU\\
    \hline
    \textbf{BoostCom} &  \textbf{BGV} &\cmark     & \cmark        & \textbf{CPU/GPU}\\
    \hline
  \end{tabular}
  }
\end{table}


Infrastructure acceleration is an approach to accelerate FHE operations with the use of hardware accelerators and efficient software implementation. It is used along with algorithmic improvement to achieve desirable performance. 
Table \ref{tab:scheme_comparison} shows the comparison of the prior works with BoostCom on infrastructure acceleration of FHE. Among all the works on infrastructure accelerations, only ours focuses on boosting the latency of comparison operations on the BGV scheme. Furthermore, the infrastructure acceleration from the prior works could be divided into two categories: operation-wise acceleration and end-to-end acceleration. For the former, the acceleration is only targeting reducing the latency of each primitive FHE operation separately such as multiplication, addition, rotation, etc. Therefore they only estimate the total execution time of an application that runs on their proposal by the latency of each FHE operation. The proposals belonging to this category typically ignore the problem of dynamic memory allocation, different levels of the ciphertext operand, noise estimation, etc. since these problems may not arise when only accelerating each of the operations separately. For the latter, the acceleration takes into account these problems and is typically used to accelerate real-world libraries such as HElib~\cite{Helib} and Microsoft SEAL~\cite{sealcrypto}. The end-to-end acceleration has a more immediate impact than operation-wise acceleration. It can be used to improve the execution time of the application that uses the real-world HE library immediately. In contrast, the operation-wise acceleration needs more work to gather them to be usable to truly run an application on it. Moreover, for both categories, the acceleration is divided by the type of hardware platform such as ASIC, FPGA, CPU (with new instructions),  GPU, and mixed CPU/GPU.


\textbf{Algorithmic acceleration.} To boost the comparison operation on BGV/BFV, some algorithmic improvements have been proposed~\cite{BGV_comparison, Tan_scheme}. The scheme results in a slightly faster speed for performing comparison compared to TFHE when the number of messages being compared is large. However, when only comparing a single message in a ciphertext, the comparison latency becomes very expensive. Typically this problem arises when the comparison is used to determine the taken branch path. Our works proposed an optimization to mitigate this problem called \emph{non-blocking comparison}. The optimization overlaps the comparison operation with other works that do not depend on the comparison result. 

\textbf{Operation-wise acceleration with GPUs}. 
The works in~\cite{TensorFHE,gme2023,HE-Booster} propose a GPGPU-based FHE acceleration solution called TensorFHE, GME, and HE-Booster, respectively. TensorFHE utilizes algorithm optimization, Number Theoretic Transform (NTT) optimization, and data layout optimization to achieve significant performance improvement for FHE arithmetic operations. \ardhi{It} also utilizes tensor cores to speed up the NTT operation. GME introduces a new NoC that connects all scratchpad memory in the GPU to reduce access to the main memory during NTT operation. HE-Booster improves the FHE arithmetic operation by improving the GPU NTT implementation from ~\cite{GPU_NTT} with fine-grain synchronization on every iteration of NTT computation.

\textbf{Operation-wise acceleration with ASIC/FPGA}. Several works in this category include ~\cite{F1, BTS, CraterLake, FxHENN, SHARP_ISCA23, ARK}. These proposals introduced an NTT unit for processing radix-2 NTT. CraterLake~\cite{CraterLake} is the first FHE accelerator to achieve high performance on unbounded FHE programs while prior accelerators are only efficient on a limited subset of simple FHE computations~\cite{F1}. CraterLake~\cite{CraterLake} is a uniprocessor with specialized functional units that span a wide vector space. The design is statically scheduled in order to take advantage of the regularity of FHE computations. SHARP~\cite{SHARP_ISCA23}, reduces the computation latency of the FHE operation by limiting the size of the prime modulus to only 36-bit. This will translate into lower memory bandwidth demand for the accelerator's memory thus improving the performance. \ardhi{Although the infrastructure acceleration using ASIC/FPGA may offer higher acceleration and better power efficiency than GPUs, they are constrained by higher development time, lower generality and lower flexibility. In contrast, CPUs/GPUs are widely available, allowing quick deployment. Their flexibility allow changing schemes, algorithms, and implementations easily.}


\textbf{End-to-end acceleration}. 
Intel HEXL introduced a new CPU instruction for processing 512-bit vectors, speeding up element-wise operations and NTT. However, this only benefits power-of-two polynomial rings, and comparison operations remain slow in such rings.
The work in~\cite{HE_on_gpu} proposed the acceleration of the BFV scheme in the Microsoft SEAL library for power-of-two polynomial rings, element-wise operation, and key-switching. Compared to our work, this work provides general FHE acceleration, while our proposal focuses on accelerating comparison operation on the BGV scheme.


\section{Bottleneck of uFHE Comparison}
\label{sec:bottleneck_analysis}

The state-of-the-art BGV comparison implementation is in HElib~\cite{BGV_comparison}. It was reported that it was up to $3\times$ faster than prior work based on BGV/BFV, and achieved even better performance than bit-wise FHE schemes in basic comparison tasks such as less-than, maximum, and minimum operations. However, each comparison still takes up to several seconds, hence we argue for the need for infrastructure acceleration. 

\begin{figure}[htbp]
  \centering 
  \includegraphics[width=0.8\linewidth]{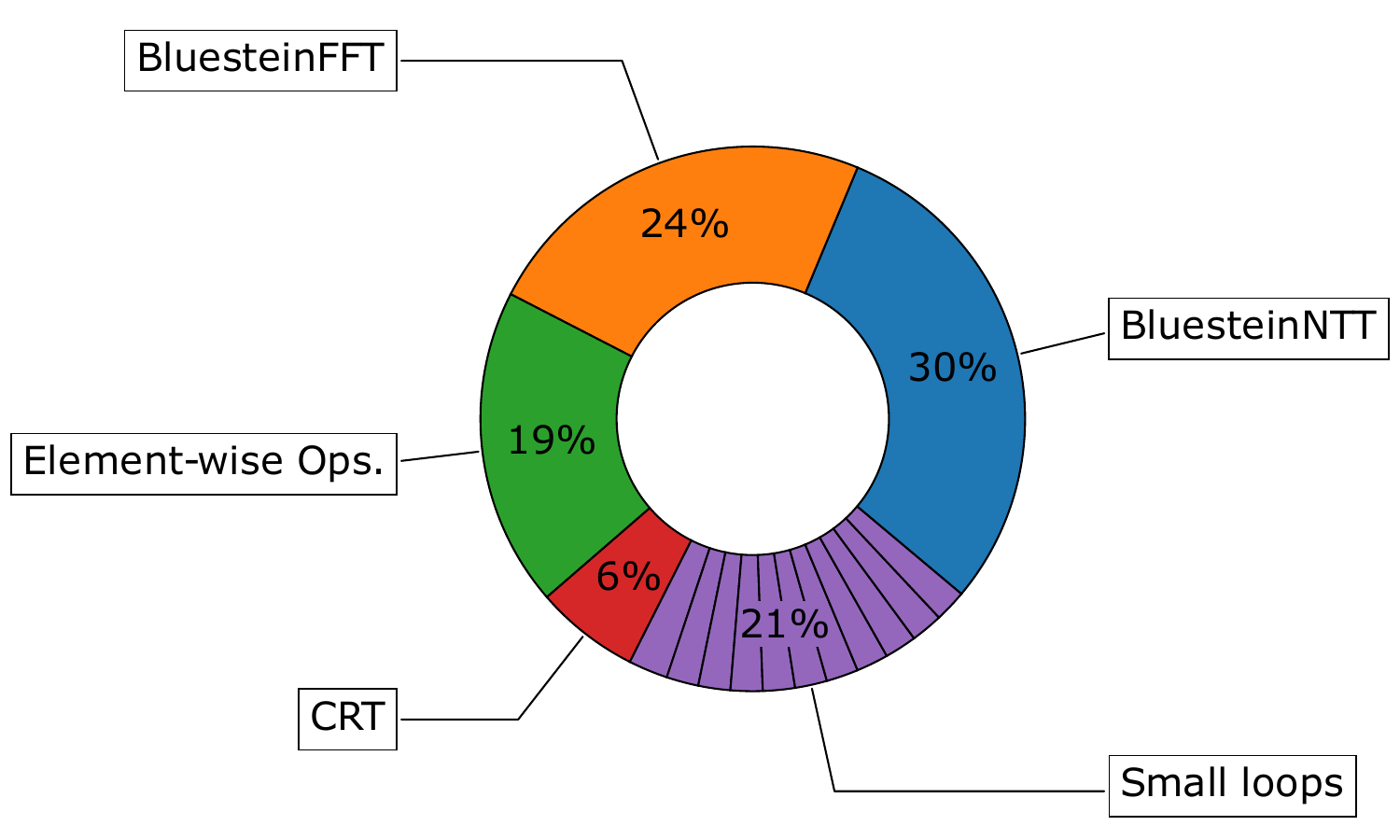}
  \vspace{-0.15in}
  \caption{Breakdown of BGV comparison time for Bivariate circuit with parameters $m=34511$, $p=3$, and $d=6$.}
  \label{fig:profile-single-bivar}
\end{figure}

To accelerate BGV comparison in HElib, we first identify the bottlenecks in the library component. To achieve this, we perform profiling and measure the execution time breakdown based on the components in the library. Figure \ref{fig:profile-single-bivar} shows the execution time breakdown of the comparison operation based on the primitive HElib components. For brevity, the figure only shows the profiling results from the Univariate case, but we note that the Bivariate case exhibits similar results. The platform we used for the profiling is detailed in Table \ref{tab:Parameters}. 

The figure shows that the execution time is mainly spent on three components: BluesteinNTT, BluesteinFFT, and Element-wise operations. Upon code inspection, we found that they are also highly parallelizable, so offloading them to GPU could be fruitful. In contrast, the "Small loops" component is also quite significant. It consists of many small loops scattered inside the library. While the code is parallelizable, the degree of the parallelism is too small to compensate for the overheads of memory copy, memory allocation, and kernel launch. Therefore,  offloading these codes to GPUs may not yield net performance improvement. For the Chinese Remainder Theorem (CRT), although the code involves multiple loops, the most time-consuming loops in this component involve the computation of a big integer and storing the final result in it.  Currently, there is no support for big integer data types on GPUs, whereas a highly optimized library for CPUs exists\cite {GMP}. Therefore, both CRT and Small loop components may not benefit from GPU offloading; instead, we will utilize CPU for their parallelization. Note that CRT parallel execution on multiple CPU cores is already the case in HElib, and we keep it that way. Furthermore, we add parallel execution of "Small loops" components on CPUs.

\section{The Design of BoostCom}
\label{sec:design}

This section describes BoostCom, our solution for BGV comparison operation acceleration through the use of GPU and multiple CPU threads. After we conduct the execution time breakdown from the previous section, we discover some primitive components inside the HElib that need to be offloaded to the GPU and what steps in the algorithm to look out for the possibility of acceleration with multi CPU threads.

\subsection{Multi-Level Heterogeneous Parallelization}

Profiling results (Section \ref{sec:bottleneck_analysis}) identified  
BluesteinNTT, BluesteinFFT, and element-wise operations in BivarLT/BivarEQ/UnivarLT+EQ as taking roughly three quarters of the execution time. Thus, an obvious acceleration step is to offload them to the GPU to benefit from the massive parallelism on the GPU. 
However, after offloading, through profiling we found that the GPU utilization is less than 10\%. This is because the parallelism granularity of the operations is insufficient to outweigh the overheads associated with memory copying, memory allocation, and kernel launching. Meanwhile, the CPU is mostly idle waiting for GPU computation results. To address both problems, we propose heterogeneous parallelization where higher-level parallelization is performed at the CPU.

\begin{figure}[htbp]
  \centering 
  \includegraphics[width=0.8\linewidth]{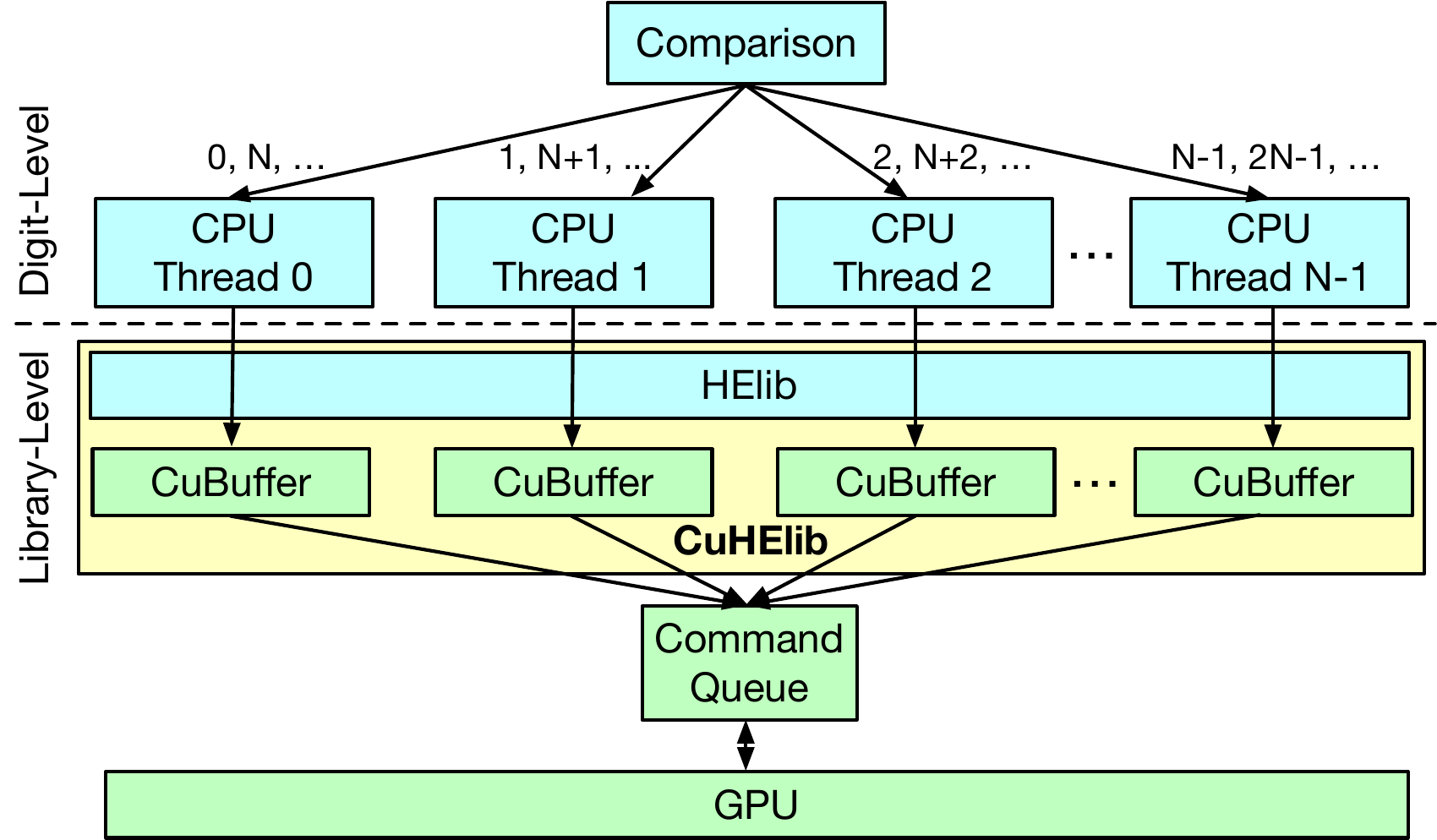}
  \vspace{-0.15in}
  \caption{Illustrating Boostcom's heterogeneous parallelism: digits are computed across multiple CPU threads, while primitive operations in each digit are offloaded to the GPU.}
  \label{fig:BoostComOverview}
\end{figure}

To perform parallelization on the CPU, one approach is to only parallelize the most time-consuming operations  (i.e., ciphertext exponentiation). However, this approach is challenging as the use of recursion creates loop-carried dependences. Moreover, the exponent of the parameters depends on $p$, which may exceed the number of CPU threads, making load balancing challenging. Hence, we explore an alternative approach of parallelizing across digits. As discussed earlier in \ref{subsec:comparison_algorithm}, the computation of each digit in $LT$ and $EQ$ has no dependence on the computation of other digits. $LT_{ij}$ computes $LT$ with digit input $a_{ij}$ and $b_{ij}$ only. The computation of each digit is also highly parallel. Hence we adopt a heterogeneous parallelism strategy, where we use GPU for specific computations for each digit in parallel, and utilize multicore CPUs to exploit digit-level parallelism. This is illustrated in Figure~\ref{fig:BoostComOverview}. 

The parallelization for digit computation is wrapped inside a library which we name cuHElib, built on top of HElib. We added multiple buffers (called CuBuffers) to hold data in the GPU memory, a command queue to dispatch tasks to the GPU, and changed the GPU task offloading strategy. The library offloads each expensive operation or function as a task (BluesteinNTT, BluesteinFFT, and Element-wise operations) to the GPUs. At the digit level, the computation of $LT$ and $EQ$ of different digits are computed across multiple CPU threads simultaneously. To avoid races and synchronization, we allocate a separate GPU buffer for each CPU thread.  

Since the CPU and GPU have separate memories, offloading computation tasks to the GPU requires copying data to a GPU buffer, launching a kernel to compute the task, and then copying the result back to the CPU. If Unified Memory (UM)~\cite{nvidiaPascal} is supported, the copying may be performed implicitly as the CPU and GPU share virtual memory address space. However, to avoid page thrashing and page faults while a kernel is running, we use explicit copying with careful timing.

\subsection{GPU-related Optimizations}
\textbf{Branch Removal for Faster BluesteinNTT.}
The BluesteinNTT computation involves element-wise multiplication (2 times), radix-2 NTT/INTT conversion, element-wise addition, and polynomial filtering. To accelerate it, we adopted the state-of-the-art radix-2 NTT/INTT implementation ~\cite{GPU_NTT}, applied an optimization~\cite{HE-Booster}, and used the Barret reduction for modular operations~\cite{SEED2022_GPUNTT}. We discovered that the remaining performance bottleneck is in polynomial filtering, which is not parallelizable due to loop-carried dependency.

Polynomial filtering alters the polynomial length from $m$ to $N$. In Listing \ref{lst:cpu_polynomial_filter}, the update of the variable $j$ is control-dependent on the loop iterator $i$, creating a loop-carried dependence that hinders loop-level parallelization. If executed sequentially with a single GPU thread, it would be inefficient due to the comparatively slower speed of a single GPU thread compared to a CPU thread. Instead, we propose a \emph{branch removal} optimization by breaking down the code into two phases (Listing \ref{lst:lockless_polynomial_filtering}): the \emph{offline phase} and the \emph{online phase}. The offline phase removes loop-carried dependences by pre-computing indices to set the target index for \emph{final\_result}. This is achieved by computing the prefix-sum of the value array of \emph{ZmStar}. Additionally, since all the inputs for index pre-computation are available before FHE computation, we can pre-compute it on the CPU. As a result, the online phase, when it performs selective copy, becomes parallelizable as we remove the branch and can benefit from GPU execution. This transformation also leverages efficient GPU pipeline computation and enables the use of multi-streaming to further improve GPU utilization.

\begin{lstlisting}[language=C++, caption=BluesteinNTT polynomial filtering code showing loop-carried dependence due the if statement and j++. ,frame=lrtb, label={lst:cpu_polynomial_filter}, mathescape]
for (i = 0, j = 0; i < m; i++)
  if (zMStar->inZmStar(i))
    final_result[j++] = coeff(result, i);
\end{lstlisting}

\begin{lstlisting}[language=C++, caption=\emph{Branch removal} optimization that removes loop-carried dependence in polynomial filtering. ,frame=lrtb, label={lst:lockless_polynomial_filtering}, mathescape]
//offline phase: index pre-comp. to remove loop-carried dependence
prefixSum(sumZmStar, inZmStar, getM);
//online phase:selective copy executed in parallel with GPU
__global__ filterBluestein(tmp, inZmStar, sumZmStar, m){
    int i = blockDim.x * blockIdx.x + threadIdx.x;
    if (i < m && inZmStar[i] != 0)
      final_result[sumZmStar[i]] = result[i]; }

\end{lstlisting}

\noindent\textbf{Plan Reuse for BluesteinFFT Acceleration.} BluesteinFFT significantly contributes to the comparison operation latency in HElib. To ensure the correctness of the ciphertext, HElib checks the noise level after each operation using BluesteinFFT. While one could use a very large $Q$ value to prevent noise budget exceedance, this may reduce computation efficiency. Opting for smaller $Q$ values, though requiring noise estimation using BluesteinFFT, may enhance computation efficiency.

HElib utilizes the CPU library PGFFT, which we replace with the cuFFT library for GPU offloading. Before using BluesteinFFT with cuFFT, a configuration step is necessary, involving plan creation for optimal thread organization. Two distinct strategies are under consideration to optimize the utilization of cuFFT: the first involves the creation of the execution plan before every BluesteinFFT operation, a straightforward yet computationally expensive approach; the second strategy, denoted as \emph{plan reuse}, configures the plan once at the initiation of FHE computation. Subsequently, during the execution of BluesteinFFT, pointers for twiddle factors and the GPU execution plan are conveyed, effectively eliminating the need for plan creation on the critical path of the operation.

\noindent\textbf{Layout Transformation for Efficient Element-Wise Ops.} Element wise operation in HElib multiplies two matrices of size  $(L+1)\times\varphi(m)$ by iteratively multiplying and adding. Each matrix is dynamically allocated because a homomorphic operation may add and/or delete rows during execution. 
The dynamic allocation may result in non-contiguous memory addresses, which creates a problem for \emph{cudamemcpy} which only copies contiguous memory address range. Thus, copying an entire matrix to the GPU using \emph{cudamemcpy} may lead to copying unrelated data. Moreover, copying the matrix result back is not feasible, as it may overwrite unrelated data processed by other threads. We explore several options to address the issue.  

One possible approach is to perform the element-wise operation row-by-row, \ardhi{which is an approach implemented for CPUs in Intel HEXL library~\cite{IntelHEXL}}. If we use this approach for GPU, we may suffer from high memory copying latencies for each row of the matrices and from a low degree of parallelism on the GPU, which may result in underutilized GPU. An alternative approach involves copying the entire matrix to the GPU row-by-row and then executing the element-wise operation for the entire matrix. This generally reduces the total kernel time. However, it still results in PCIe bandwidth wastage since only a small amount of data is copied to the GPU multiple times. 

\begin{figure}
  \centering 
  \includegraphics[width=0.70\linewidth]{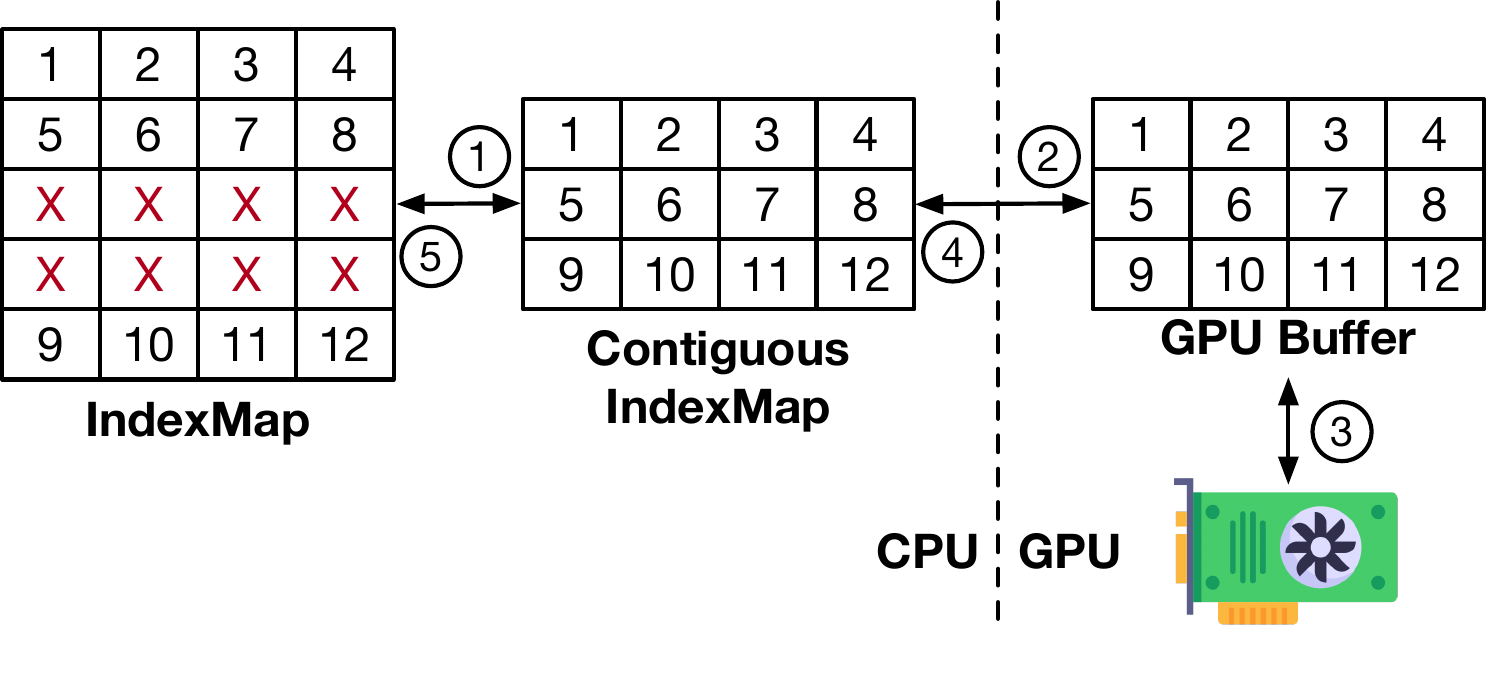}
  \vspace{-0.15in}
  \caption{Layout optimization for offloading the element-wise operation to the GPU, utilizing additional copying at the CPU side to maximize the CPU-GPU \emph{memcpy} bandwidth and parallelization degree.}
  \label{fig:contiguousbuffer}
\end{figure}

Thus, we use a third approach, which we refer to as \emph{layout transformation}, to create a contiguous memory allocation in the CPU buffer, which allows the element-wise operation for the entire matrix to be offloaded in a single GPU kernel. Figure \ref{fig:contiguousbuffer} illustrates the steps for this optimization: \circled{1} data from the original matrix with non-contiguous row locations is copied over to a new buffer with contiguous mapping. \circled{2} the entire matrix with contiguous rows is transferred to the GPU. \circled{3} element-wise operation is performed on the GPU, producing results in the GPU buffer. \circled{4} data in the GPU buffer is copied back. \circled{5} the resulting matrix is copied back to the original buffer. This approach incurs additional CPU-to-CPU memory copying but maximizes PCIe bandwidth utilization and allows a high degree of GPU  parallelization. 

\subsection{Algorithm-level Optimizations}

\textbf{Slot Compaction.} SIMD-style processing facilitates the simultaneous manipulation of tens of thousands of numbers placed in slots and encoded within a single ciphertext, whereby an operation on the ciphertext is performed on all numbers. A high slot utilization increases both compute and memory efficiency. However, our analysis reveals that slot utilization is often low, especially for comparison, for three reasons. First, there are often discrepancies between the input size alignment and the available ciphertext slots, which persist even after optimizations. For instance, AlexNet's input size is $224\times 224$, while SEAL~\cite{sealcrypto} supports a maximum of 16,384 slots per ciphertext. Consequently, the input is partitioned into $\lceil\frac{224\times 224}{16384}\rceil=4$ ciphertexts, resulting in 24\% of slots being unused.

\begin{figure}[]
  \centering 
  \includegraphics[width=\linewidth]{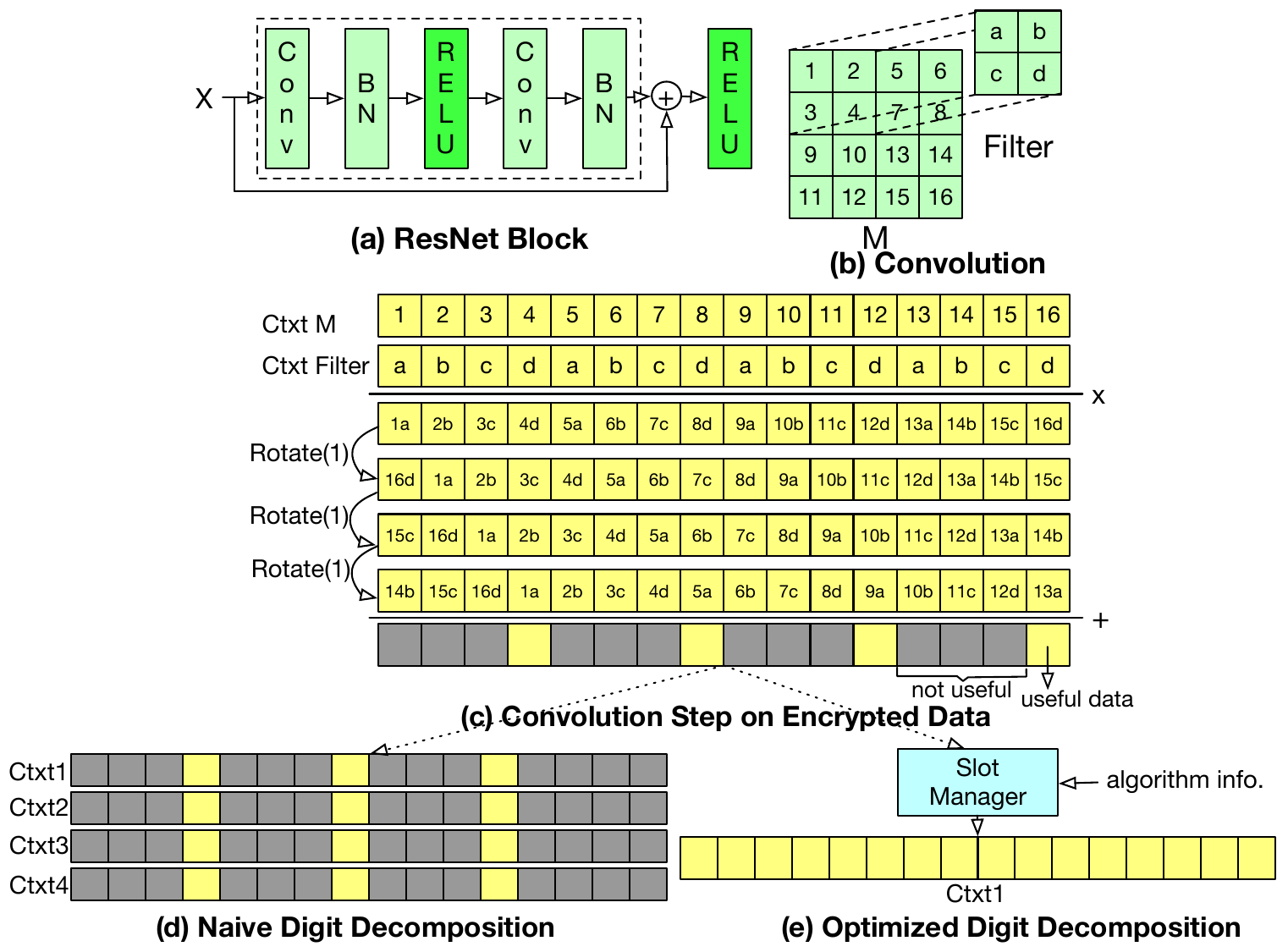}
  \caption{Illustration of: (a) ResNet block containing convolution (Conv), batch normalization (BN), and ReLU; (b) Convolution filter; (c) Convolution steps on encrypted data resulting in unused slots; (d) Naive digit decomposition with many unused slots; and (e) Optimized digit decomposition with slot compaction.}
  \label{fig:conv}
\end{figure}

The second reason for slot under-utilization is slot wastage produced by prior operations. For example, in machine learning workloads, comparison occurs after other operations, such as matrix multiplication or convolution, that produce the waste. The third reason is that standard optimizations, e.g. in Helayers\cite{helayers},  to reduce future operations (by duplicating numbers in different slots) can actually amplify the slot wastage in the comparison operation. These create an opportunity to perform {\em slot compaction} prior to performing the comparison. Figure~\ref{fig:conv} illustrates an example of a neural network where convolution (and batch normalization) precedes ReLU in which comparison is performed (part (a)). For convolution between a matrix M and a filter (part (b)), the matrix M fills up all 16 ciphertext slots. Then, the convolution filter is duplicated to fill up slots (part (c)), in order to reduce the number of future multiplications, rotations, and additions, and to improve slot utilization. To obtain the convolution results, the multiplication is followed by only three sets of rotate-and-accumulate. The convolution results occupy the 4th, 8th, 12th, and 16th slots (shown in yellow),  while all other slots do not contain useful values (or wasted). 

Next, to achieve comparison, Figure \ref{fig:conv} (part (d)) illustrates the state-of-the-art practice where each convolution result is decomposed into digits (four digits are illustrated), resulting in amplifying the slot wastage across four ciphertexts, where even useless values are also decomposed into four digits. For comparison, only 25\% slots have useful digits that are needed, which presents an opportunity for compaction.  Our approach is shown in part (e), where we consolidate digits from all numbers into a single ciphertext. Through slot compaction, the comparison can now work on fewer ciphertext inputs, substantially reducing memory usage and unnecessary computation. 

Realizing slot compaction in the Helayer is difficult because it cannot distinguish   slots which contain useful data vs. those who do not, hence it must conservatively assume that all slots are useful. Besides, the existence of non-useful slots arises only when comparison is preceded by certain operations like convolution or matrix multiplication, so the Helayer cannot identify slot usefulness without algorithmic information. To overcome this challenge, we design a slot manager (SM) that preserves  algorithm information to track slot usefulness to guide slot compaction after digit decomposition. With such information, SM can minimize memory usage by distributing digit decomposition across as few ciphertexts as possible. 

When comparison is not preceded by other operations, we just perform slot compaction for the case when the input size does not align with the ciphertext format. 


\begin{minipage}[htbp]{0.4\linewidth}
\begin{lstlisting}[language=C++, caption=Private query on plaintext data.,frame=lrtb, label={lst:privateQueryDatabase}, mathescape]
privateQuery(q, op1, op2){
 if(q == add)
    Data += op1
 else if(q == mult)
    Data *= op1
 else if(q == power)
    Data = Data^op2
 else
    Data = Data }
\end{lstlisting}
\end{minipage}
~
\begin{minipage}[htbp]{0.5\linewidth}
\begin{lstlisting}[language=C++, caption=Private query on encrypted data., frame=lrtb, label={lst:privateQueryDatabaseEncrypted}, xleftmargin=0.5cm]
privateQuery(q, op1, op2){
 c1 = EQ(q, add)
 c2 = EQ(q, mult)
 c3 = EQ(q, pwr)
 
 Data1 = Data + op1
 Data2 = Data * op1
 Data3 = Data.Power(op2)
 Data = Data1 * c1 + Data2 * 
        c2 + Data3 * c3 }
\end{lstlisting}
\end{minipage}

\begin{minipage}[htbp]{0.90\linewidth}
\begin{lstlisting}[language=C++, caption=Private query with non-blocking comparison.,frame=lrtb, label={lst:privateQueryDatabaseEncryptedOptimized}, xleftmargin=0.2cm]
EvalBranch(c1, c2, c3, q){
 c1 = EQ(q, add)
 c2 = EQ(q, mult)
 c3 = EQ(q, pwr)
}
privateQuery(q, op1, op2) { 
 helper_thread(EvalBranch(c1, c2, c3, q))
 Data1 = Data + op1
 Data2 = Data * op1
 Data3 = Data.Power(op2)
 thread_1.join()
 Data = Data1 * c1 + Data2 * c2 + Data3 * c3 }
\end{lstlisting}
\end{minipage}

\noindent\textbf{Non-Blocking Comparison.}
When many numbers are compared together, the cost of comparison operation could be amortized using SIMD-style processing. However, when an application only needs to compare a pair of numbers (or a small number of pairs), comparison latency is hard to amortize. This case occurs when the comparison occurs inside an \emph{if} statement. 

Listing~\ref{lst:privateQueryDatabase} shows an example code that performs a query without FHE (i.e., on unencrypted data). It takes three inputs: query type (\texttt{q}) and two data operands (\texttt{op1} and \texttt{op2}). The code performs an operation (addition, multiplication, or exponentiation) based on the query type, with operand value specified by one of the two data operands. It has three comparisons each involving a pair of numbers. The semantic-equivalent FHE version is shown in Listing~\ref{lst:privateQueryDatabaseEncrypted}. With FHE, the query type is not in plaintext form, hence we must use the EQ(.) function to test for equality. Furthermore, the comparison results are also in ciphertext, hence conditional branches are replaced by code straightlining, resulting in  Listing~\ref{lst:privateQueryDatabaseEncrypted}. 

To hide the comparison latency that is hard to amortize, we propose \emph{non-blocking comparison}. When the comparison is solely used to determine the taken branch path, there is no dependency relation between the main computation and the branch evaluation. Consequently, we can execute the branch evaluation and the main computation concurrently. Listing~\ref{lst:privateQueryDatabaseEncryptedOptimized} shows the resulting code with our \emph{non-blocking comparison} optimization. The branch evaluation that computes equality functions EQ(.) is performed by a helper thread in parallel to the arithmetic operations performed by the main thread. The final data update is performed after the helper thread joins the main thread.

 \begin{figure}[htbp]
  \centering 
  \includegraphics[scale=0.4]{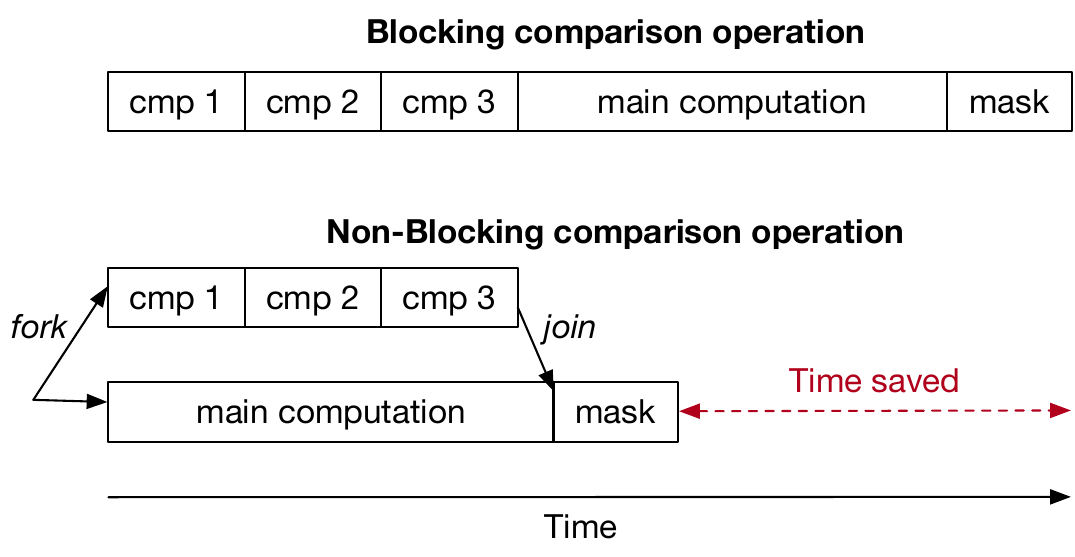}
  \vspace{-0.2in}
  \caption{Illustrating the saved cycles due to the non-blocking comparison optimization.}
  \label{fig:nonblocking_comparison}
\end{figure}

To illustrate the benefit, Figure \ref{fig:nonblocking_comparison} compares the original straightlined code performance (top) vs. with our non-blocking optimization (bottom). With non-blocking, the execution of branch evaluation overlaps with the main computation.

\noindent\textbf{Non-Blocking Comparison.} The conventional wisdom in BGV parameter selection favors {\em power-of-two} (PoT) polynomial ring degrees. However, non-PoT degrees, achieved by choosing prime numbers or cyclotomic polynomial ring orders, offer enhanced performance by ensuring cyclic slot permutation groups~\cite{BGV_comparison, GentryEuroCrypt}. While prior works often overlooked the impact of PoT degrees on comparison performance, our experimentation results demonstrate that non-PoT degrees lead to significantly faster comparisons.

 \begin{figure}[htbp]
  \centering 
  \includegraphics[width=0.9\linewidth]{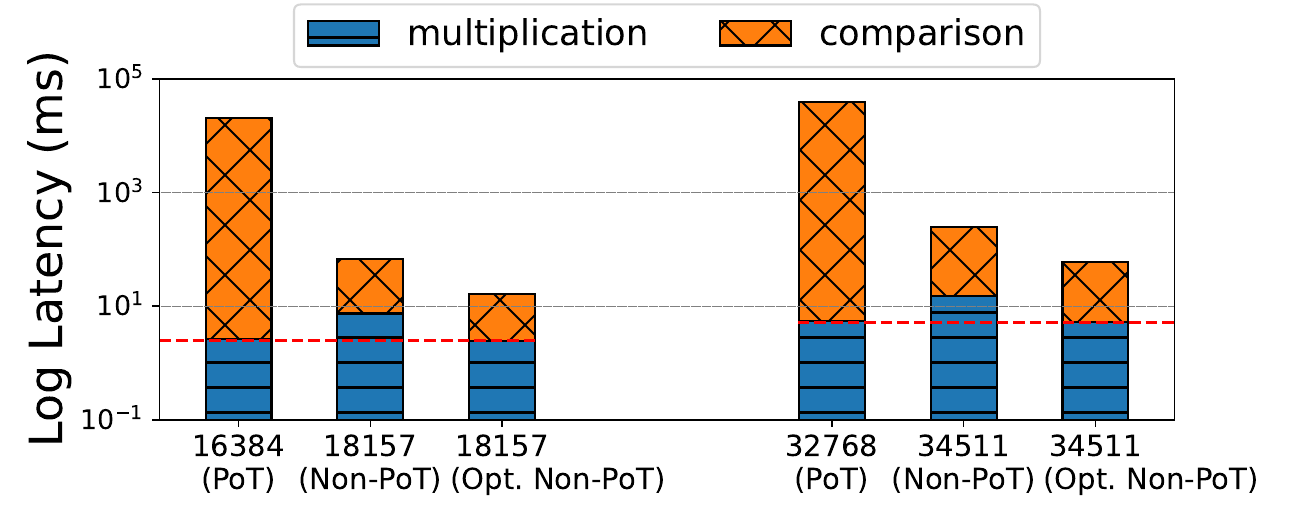}
  \vspace{-0.15in}
  \caption{Comparison of the homomorphic operation latencies of using the power of two vs. non-power of two vs. optimized non-power of two polynomial rings.}
  \label{fig:pot_nonpot}
\end{figure}

\begin{table}[htbp]
\small	
  \caption{Parameters and Statistics}
  \vspace{-0.15in}
  \label{tab:Parameters}
\centering
  {
 \scriptsize
  \begin{tabular}{|c||c|c|c|c||c|c|}
    \hline
    Params & ($p$ $m$ $N$) & Circuit & ($d$ $l$) & log(Q) & $\lambda$ & no of int \\
    \hline \hline 
     \multirow{2}{*}{p1} &\multirow{2}{*}{(3 34511 34510)}& B & (6 7) & 324 & 298 & 290\\ 
                                                                         & & U & (16 4) & 472 & 189 & 507\\
    \hline
    \multirow{2}{*}{p2} &\multirow{2}{*}{(5 19531 19530)} & B  & (7 4) & 324 & 155 & 697\\ 
                                                                         & & U & (7 6) & 354 & 141 & 465\\ 
    \hline
    \multirow{2}{*}{p3} &\multirow{2}{*}{(7 20197 19116)} & B  &  (6 4) & 354 & 137 & 531\\ 
                                                                         & & U & (8 4) & 406 & 110 & 531\\ 
    \hline
    \multirow{2}{*}{p4} &\multirow{2}{*}{(11 15797 15796)} & B  & (5 4) & 342 & 162 & 359\\ 
                                                                         & & U &(5 5)  & 378 & 145 & 287\\ 
    \hline
    \multirow{2}{*}{p5} &\multirow{2}{*}{(13 30941 30940)} & B  &(5 4)  & 354 & 338 & 1547\\ 
                                                                         & & U &(4 6)  & 378 & 313 & 1031\\ 
    \hline
    \multirow{2}{*}{p6} &\multirow{2}{*}{(17 41761 41760)} & B  &(4 4)  & 413 & 402 & 1305\\ 
                                                                         & & U &(7 3)  & 472 & 344 & 1740\\ 
    \hline
    \multirow{2}{*}{p7} &\multirow{2}{*}{(19 29989 29988)} & B  &(4 4)  & 378 & 302 & 833\\ 
                                                                         & & U &(5 4)  & 385 & 296 & 833\\ 
    \hline
    \multirow{2}{*}{p8} &\multirow{2}{*}{(23 37745 30192)} & B  &(5 3)  & 413 & 275 & 838\\ 
                                                                         & & U &(9 2)  & 456 & 245 & 1258\\ 
    \hline
    \multirow{2}{*}{p9} &\multirow{2}{*}{(29 18157 17820)} & B  &(5 3)  & 360 & 175 & 990\\ 
                                                                         & & U & (6 3) & 413 & 150 & 990\\ 
    \hline
    \multirow{2}{*}{p10} &\multirow{2}{*}{(31 52053 34700)} & B  &(5 3)  & 512 & 252 & 2313\\ 
                                                                         & & U & (4 4)  & 512 & 252 & 1735\\     
    \hline
  \end{tabular}}
\end{table}

As a demonstration, Figure \ref{fig:pot_nonpot} presents stacked bars representing the latencies of a single multiplication and a single comparison in {\em logarithmic scale} for a pair of $m$ values. To qualify this, while ensuring a security level of $\lambda > 128$ bits, we have chosen two non-PoT $m$ values, 18,157 and 34,511, to correspond with two specific PoT $m$ values, 16,384 and 32,768. To ensure a meaningful comparison, these selections have been designed such that the non-PoT $m$ values yield a slightly larger count of SIMD slots and an enhanced security level, as guided by the recommendations from~\cite{albrecht2015, BGV_comparison}. For each $m$ value, we show the stacked latencies for three cases: PoT, unoptimized non-PoT, and non-PoT with our optimizations. 

The figure shows that after optimization non-PoT latencies are over two orders of magnitude faster. Therefore, in this paper we choose non-PoT polynomial ring order due to the computation efficiency.

\section{Methodology}

\paragraph{Experiment Platforms}
We evaluate BoostCom on a combination of GPU and CPU platforms. The GPU platform has an NVIDIA RTX 3090 GPU with 82 Streaming Multiprocessors (SMs). Each SM contains 128 CUDA cores, operating at a core clock speed of 1695 MHz. The GPU is equipped with a combined 10 MB of L1 data cache and shared memory, along with a separate 6 MB L2 cache. The GPU memory system has a 24GB size and 936 GB/s bandwidth. 

The CPU platform has an AMD Ryzen PRO 3955WX CPU with 16 cores and 128 GB of memory. Each core has a clock speed of 3.9 GHz, with a 4.3 GHz maximum turbo frequency. It has 64MB L3 Cache Memory and its main memory has eight-channel ECC DDR4-3200 DRAMs. The CPU runs Ubuntu OS version 22.04 and NVIDIA driver version 525.85.12. We used CUDA version 12.0 and GCC version 7.5.0 for compilation.

\paragraph{Workload Evaluation Methodology}
We evaluate BoostCom with full applications to measure overall application performance as well as with microbenchmark to measure comparison performance specifically. In both cases, our evaluation measures end-to-end performance, in contrast to extrapolating from the measurement of each operation that is common in prior works. End-to-end performance measurement gives a fuller and more reliable picture of the performance. For all measurements, we repeat each experiment 10 times and report their average. We use the NVIDIA Nsight system to collect hardware performance statistics.

\paragraph{State-of-the-art BGV accelerator} 
We compare our work with \ardhi{state-of-the-art FHE GPU-acceleration} HE-Booster \cite{HE-Booster}. HE-Booster accelerates NTT operations by introducing fine granularity of thread synchronization for every iteration inside NTT operations. Additionally, the paper proposes the fusion of operations within the key-switching procedure. We implemented HE-Booster in the HELib library and evaluated it end-to-end.

\paragraph{Microbenchmark} To measure comparison-only performance, we form a microbenchmark that performs a comparison of a pair of 64-bit integers. We vary the BGV parameters to form 10 different configurations following prior work~\cite{BGV_comparison}. Each configuration is expressed as a tuple of ($p \; m \; N$) and was selected to maximize the number of SIMD slots as shown in Table \ref{tab:Parameters}. They are sorted in the order of increasing plaintext modulus $p$ values. Each configuration uses bivariate and univariate circuits with differing vector space dimension $d$, vector length $l$, and the product of prime moduli $Q$. The resulting security level $\lambda$ and number of integers that can fit in one ciphertext are shown in the last two columns.  

\paragraph{Applications}
\ardhi{As there is currently no standardized benchmark for evaluating comparison operations in BGV, we developed \textbf{mlp}, \textbf{img\_col}, and \textbf{private\_q}, and adopted  \textbf{sorting} and \textbf{min} from prior work~\cite{BGV_comparison}. Below, we provide details for each benchmark:}

\textbf{sorting} is an application that sorts an array of 16 encrypted 32-bit integers from~\cite{BGV_comparison}. It uses univariate circuit for comparison, utilizes a matrix of Hamming weights to establish the relationship between any pair of elements in the encrypted array. 

\textbf{min} is an application that finds a minimum integer from an array of 16 elements of 32-bit integers from \cite{BGV_comparison}. It uses univariate circuit and combines the Hamming weight matrix and the tournament methodology, reducing the circuit's depth for improved efficiency.

\textbf{mlp} is a simple machine learning program utilizing Multi-Layer Perceptron that we wrote to classify images. It has three layers: a fully-connected layer, ReLU, and another fully-connected layer. 
The bivariate circuit is used for comparison in the ReLU layer. mlp performs inference using encrypted 16-bit integers. The input image has 28x28 pixels,  stored in a single ciphertext. It trains on MNIST datasets and outputs ten nodes. 

\textbf{img\_col} is an image recolorizing application that we developed to calculate the distance of every pixel inside an image to a threshold value. When the distance is below the threshold, it transforms the pixel by multiplying its color value with the pre-set value. The bivariate circuit is used for comparison. This application enables private medical data image analysis on an untrusted cloud server. The input is an encrypted image, threshold value, and pre-set pixel transformation value. The input image is encoded into 16 ciphertexts and each ciphertext consists of 700 pixels.

\textbf{private\_q} is a simple application that we developed to perform a private query to manipulate data in encrypted databases, based on Listing~\ref{fig:nonblocking_comparison}. The database consists of 100 ciphertexts, and each ciphertext holds 2124 integers. This application helps evaluate the proposed \emph{non-blocking comparison}.
\section{Evaluation Results}

\subsection{Workloads Speedup}

\begin{figure}[htbp]
  \centering 
  \includegraphics[width=1.0\linewidth]{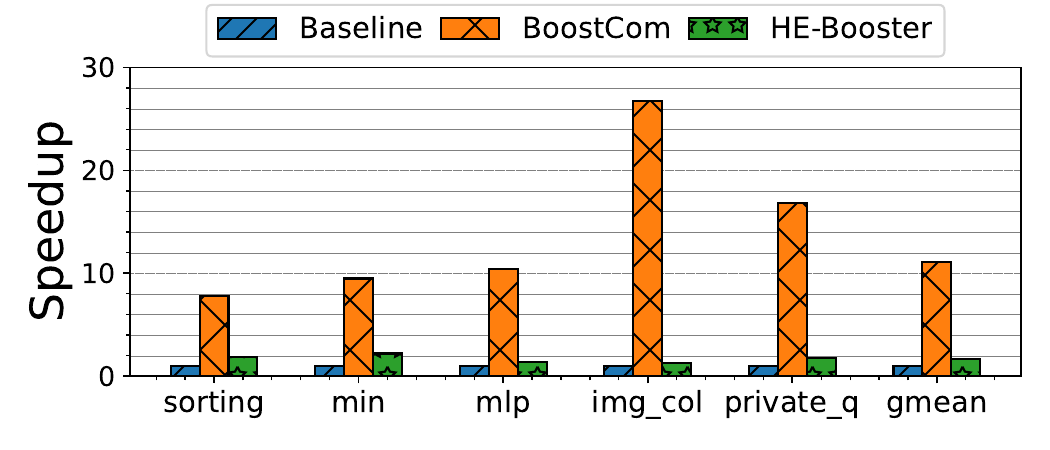}
  \vspace{-0.4in}
  \caption{The acceleration achieved by BoostCom in comparison to the baseline for five important workloads.}
  \label{fig:workload_speedup}
\end{figure}

Figure \ref{fig:workload_speedup} illustrates the speedups achieved by BoostCom and HE-booster compared to a 16-core CPU-only baseline (i.e., HElib~\cite{Helib}), for all applications and their geometric mean speedup. BoostCom (second bars) include library-level optimizations, i.e., heterogeneous multi-CPU/GPU parallelization, slot compaction, and non-blocking optimization.

\begin{table}[htbp]
\small	
  \caption{GPU Time Utilization.}
  \vspace{-0.15in}
  \label{tab:gpu_time_util}
\centering
  {
  \footnotesize
  \begin{tabular}{|p{1.4cm}||c|c|c|c|c|c|}
    \hline
     Scheme &sorting & min & mlp & img\_col & private\_q\\
    \hline \hline 
    GPU-only & 17\% & 8\% & 20\% & 15\% & 17\%\\
    \hline
    Heterogeneous & 35\% & 16\% & 41\% & 32\% & 33\%\\
    \hline
  \end{tabular}}
\end{table}

\begin{table}[htbp]
\small	
  \caption{Memory Usage for each Workload (GB).}
  \vspace{-0.15in}
  \label{tab:memoryusage}
\centering
  {
  \footnotesize
  \begin{tabular}{|p{1.85cm}||c|c|c|c|c|c|}
    \hline
     Scheme &sorting & min & mlp & img\_col & private\_q & gmean\\
    \hline \hline 
    Heterogeneous & 5.5 & 8.4 & 1.6 & 2.8 & 2.2 & 3.5\\
    Heterogeneous+SM & 4.5 & 6.5 & 1.1 & 1.6 & 1.4 & 2.5\\
    \hline
    Mem. Reduction & 19\% & 23\% & 32\% & 44\% & 35\% & 29.3\%\\
    \hline
  \end{tabular}}
\end{table}

It is important to note that the speedups are measured for \emph{end-to-end execution times}, encompassing various operations, not just the comparison. This includes all overhead such as CPU-GPU memory copy, kernel launches, synchronization, etc. As shown from the figure, BoostCom achieved a speedup of \(11.1\times\) (up to $26.7\times$). In contrast, the state-of-the-art HE-Booster only achieves an average of \(1.7\times\) speedup. Thus, our Boostcom scheme achieves a 553\% higher speedup than HE-Booster on average. 

\begin{figure}[htbp]
  \centering 
  \includegraphics[width=0.8\linewidth]{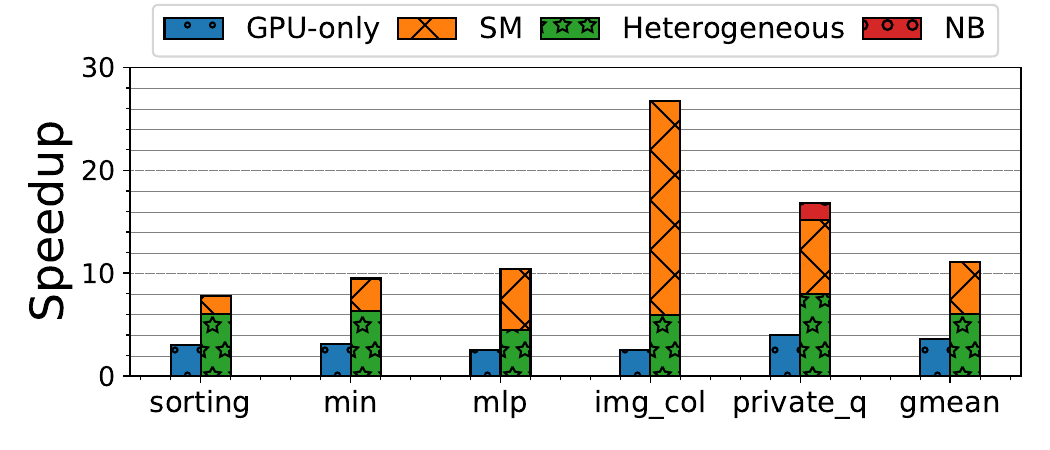}
  \vspace{-0.22in}
  \caption{The breakdown of speedup for each optimization compared to the baseline for five critical workloads.}
  \label{fig:workload_speedup_breakdown}
\end{figure}

Figure \ref{fig:workload_speedup_breakdown} illustrates the impact of each BoostCom's optimization. The GPU-only denotes library-level optimizations with offloading the computation-intensive portions of the library to the GPU with only one CPU core, SM denotes the usage of slot manager for compaction, The Heterogeneous denotes the usage of a multicore CPU to submit more work to the GPU, and NB adds the non-blocking optimization. On average, the GPU-only acceleration only achieved a gmean speedup of \(3.6\times\) over the baseline. The gmean speedup triples when multi CPU core and slot compaction is added, reaching \(11.1\times\). This demonstrates the effectiveness of Boostcom's  heterogeneous scheme and the slot compaction that reduces the number of ciphertexts involved in comparison. As depicted in the figure, each optimization demonstrates a significant effect on the speedup, highlighting the effectiveness of each of the optimization.

Boostcom's multi-level heterogeneous parallelization strategy roughly doubles the GPU utilization (Table \ref{tab:gpu_time_util}) while slot compaction reduces the memory usage by 29.3\% on average (Table \ref{tab:memoryusage}). The reduction clearly correlates with the speedups; the greater the memory usage reduction, the higher the speedup.


In the subsequent subsection, we analyze the effect of each optimization at the library level employed in BoostCom concerning only the comparison operation.

\subsection{Comparison Operation Speedup}
 \begin{figure}
  \centering 
  \includegraphics[width=0.95\linewidth]{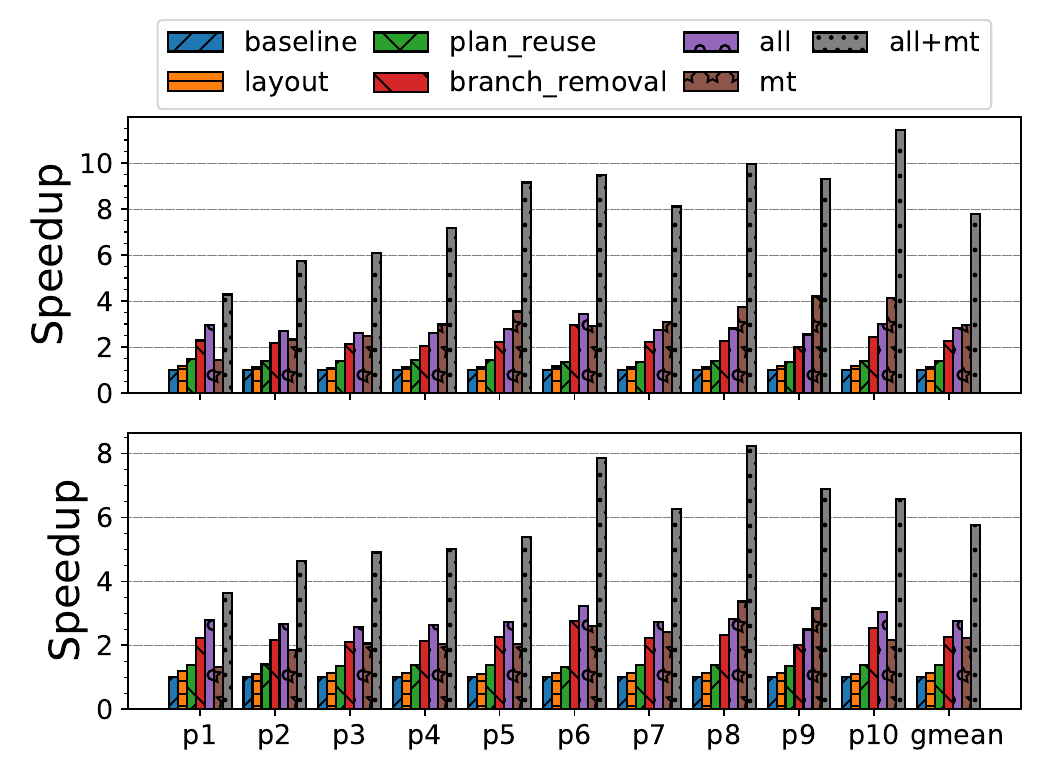}
  \vspace{-0.2in}
  \caption{Speedups of the comparison ops for the Bivariate (top) and Univariate (bottom) circuit over the baseline.}
  \label{fig:comparison_speedup}
\end{figure}

 \begin{figure}
  \centering 
  \includegraphics[width=0.70\linewidth]{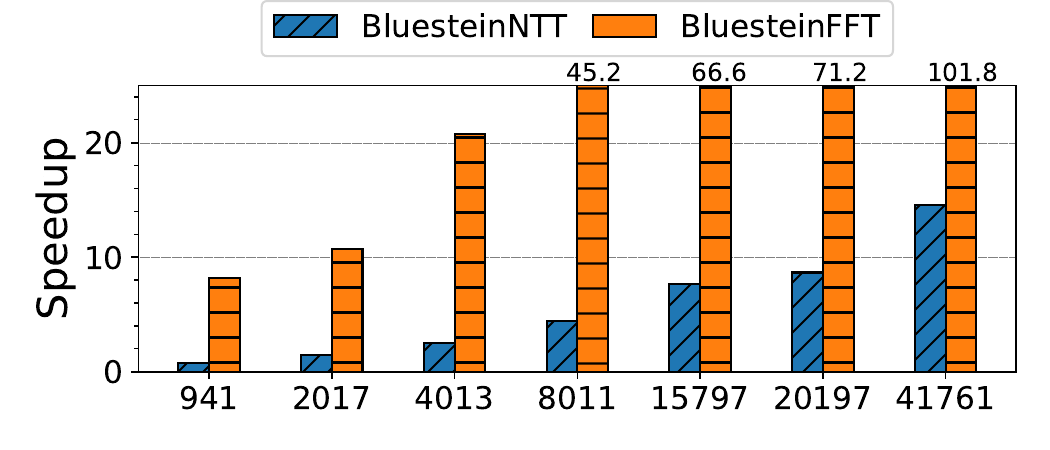}
    \vspace{-0.2in}
  \caption{The comparison between BluesteinNTT and BluesteinFFT speedup over each baseline with the increasing parameter $m$.}
  \label{fig:bluesteinSensitivityStudy}
\end{figure}

Figure \ref{fig:comparison_speedup} compares the end-to-end execution time of comparison of encrypted 64-bit int, over the 16-core CPU-only baseline for Bivariate (top) and Univariate (bottom) circuits, across 10 different BGV configurations from Table~\ref{tab:Parameters}. For each configuration, six bars are shown with increasing optimization levels, starting from the baseline, layout transformation, branch removal, plan reuse, the combination of three said optimizations   (\textit{all}), digit level parallelization with CPU multithreading (\textit{mt}), and all optimizations including multithreading (\textit{all+mt}). Note that slot compaction and non-blocking comparison optimizations are not applicable here since there is no other computation aside from the comparison itself. 

For both circuits across all configurations, each optimization adds additional speedups, indicating their effectiveness. With {\em all}, the geometric mean (gmean) speedup is 2.8$\times$ for both circuits. On its own, multithreading for digit-level parallelization is somewhat effective (gmean speedup of 2.9$\times$ (Bivariate) and $2.2\times$ (Univariate)). But when combined with all other optimizations, multithreading enables much higher speedups, reaching 7.8$\times$ (Bivariate) and 5.8$\times$ (Univariate), due to the synergistic effect where multithreading significantly improving the GPU utilization (by between 30\% and 260\%). 

Roughly, as $p$ increases, the effectiveness of multithreading increases whereas that of other optimizations remains unchanged. This is because as the degree $d$ increases, the fraction of execution time spent on the \emph{BivarCircuit}, \emph{EqualityCircuit}, and \emph{UnivarCircuit} increases. 



\subsection{BluesteinNTT and BluesteinFFT Sensitivity Study}

\paragraph{Impact of the parameter $m$} 
Increasing multiplicative depth without sacrificing security may lead to larger $m$. To evaluate its effect on BoostCom, we vary $m$ from 941 to 41,761, resulting in polynomial size expansions ranging from 2,048 to 131,072. The resulting speedups of BluesteinNTT and BlusteinFFT, calculated over CPU-only execution are shown in Figure \ref{fig:bluesteinSensitivityStudy} (top). The figure shows that the larger the $m$, the higher the speedups, indicating BoostCom's scalability. 

 \begin{figure}
  \centering 
  \includegraphics[width=0.8\linewidth]{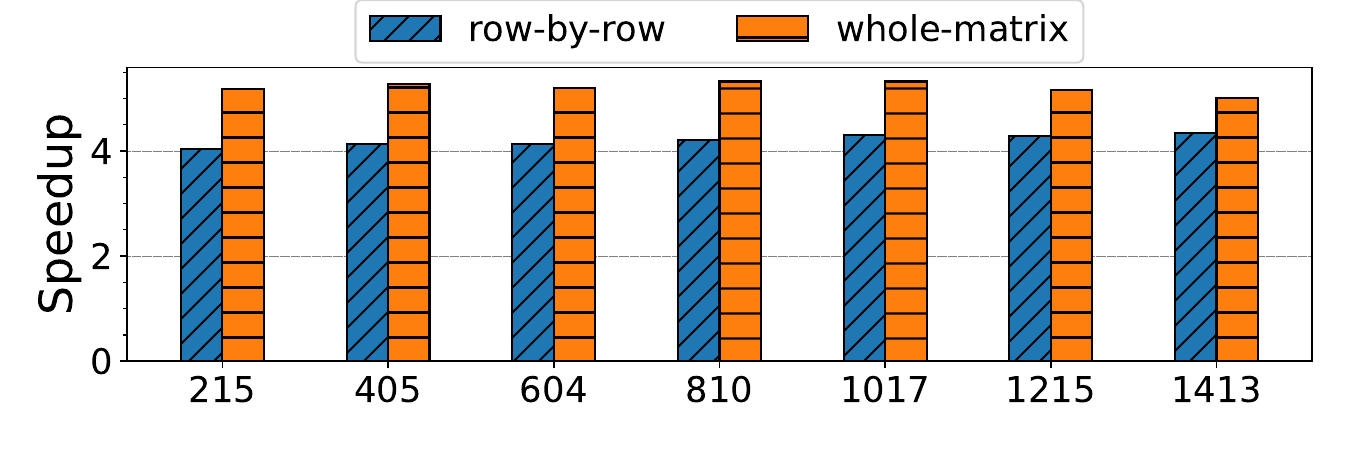}
    \vspace{-0.2in}
  \caption{Element-wise ops speedups of whole matrix approach vs. row-by-row GPU offloading, as $log(Q)$ increases.}
  \label{fig:elementwiseSensitivityStudy}
\end{figure}

 \begin{figure}
  \centering 
  \includegraphics[width=0.8\linewidth]{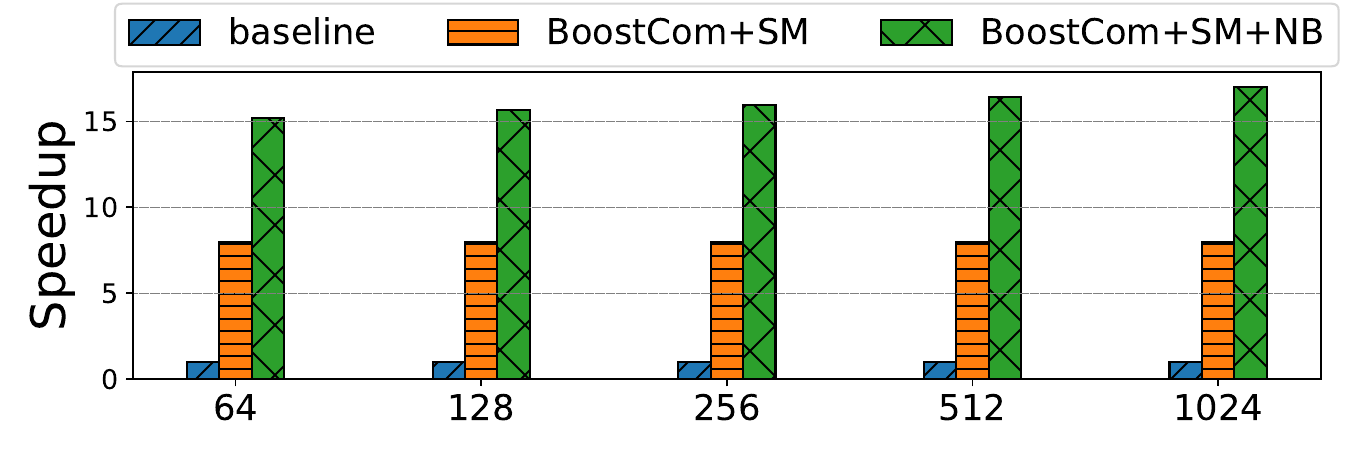}
    \vspace{-0.2in}
  \caption{Speedups of optimizations without vs. with non-blocking as the branch evaluation computation increases with larger exponent values.}
  \label{fig:privateQuery_speedup}
\end{figure}

\subsection{Element-wise Sensitivity Study}

Figure~\ref{fig:elementwiseSensitivityStudy} compares the speedups of BoostCom's \emph{layout transformation} compared to performing element-wise operation row-by-row, as $Q$ increases. 
A larger $Q$ increases the noise budget and allows a more complex application but with slower computation. The figure shows that the speedups of our layout optimization is quite stable across all values of $Q$. 

\subsection{Non-blocking Comparison Sensitivity Study}
To evaluate the sensitivity of BoostCom's \emph{non-blocking} optimization performance, we vary the exponent (\texttt{op2}) from 64 to 1024 as exponentiation is the most expensive operation.  (Figure \ref{fig:privateQuery_speedup}). The figure shows the speedups are stable, with increasing non-blocking effectiveness (as a larger portion of the branch evaluation is hidden).

\section{Conclusion}

We proposed accelerating uFHE-based BGV scheme's non-arithmetic comparisons on CPU/GPU systems through innovative optimizations, including multi-level heterogeneous parallelization, GPU optimizations, and algorithmic designs. This combination of optimizations proved highly effective, achieving an 11.1$\times$ speedup (with peaks up to 26.7$\times$) across five key FHE applications, significantly outperforming the prior approach.

\bibliographystyle{ACM-Reference-Format}
\bibliography{software,sample-base}

\end{document}